%% file: main.tex
\documentclass[prl,aps,twocolumn,preprintnumbers,amssymb,nobibnotes,nofootinbib,raggedbottom,10pt]{revtex4-1}
\usepackage{hyperref,graphicx,array,multirow,color,amsmath,mathrsfs,titlesec,shuffle,tikz}
\input{header_data.tex}

\newcommand{\la}{\langle}
\newcommand{\ra}{\rangle}

\newcommand{\Ddots}{\hbox to 1em{.\hss.\hss.\hss}}
\newcommand{\ang}[1]{\langle #1\rangle}
\newcommand{\sq}[1]{\left[ #1\right]}
\allowdisplaybreaks

\begin{document}
\title{Gravity Amplitudes From Double Bonus Relations}
\author{Shruti Paranjape$^\spadesuit$}
\author{Jaroslav Trnka$^{\spadesuit\Diamond}$}

\affiliation{$^\spadesuit$Center for Quantum Mathematics and Physics (QMAP), University of California, Davis, CA, USA}
\affiliation{$^\Diamond$Institute for Particle and Nuclear Physics, Charles University in Prague, Czech Republic}

\begin{abstract} 
In this letter we derive new expressions for tree-level graviton amplitudes in $\mathcal{N}=8$ supergravity from BCFW recursion relations combined with new types of bonus relations. These bonus relations go beyond the famous $1/z^2$ behavior under a large BCFW shift, and use knowledge about certain zeroes of graviton amplitudes in collinear kinematics. This extra knowledge can be used in the context of global residue theorems by writing the amplitude in a special form using canonical building blocks. In the NMHV case these building blocks are dressed one-loop leading singularities, the same objects that appear in the expansion of Yang-Mills amplitudes, where each term corresponds to an $R$-invariant. Unlike other approaches, our formula is not an expansion in terms of cyclic objects and does not manifest color-kinematics duality, but rather preserves the permutational symmetry of its building blocks. We also comment on the possible connection to Grassmannian geometry and give some non-trivial evidence of such structure for graviton amplitudes.
\vspace{-10pt}
\end{abstract}

\maketitle

\vspace{-0.3cm}

\section{Introduction}

\vspace{-0.3cm}

In last two decades, the study of gravitational amplitudes has been a very active area of research, leading to major discoveries and great improvement in our theoretical understanding and computational abilities. However, some major mysteries remain unresolved even for tree-level amplitudes. For example, the calculation of higher-point amplitudes using Feynman diagrams is notoriously difficult due to the presence of vertices of any multiplicity and their complicated Feynman rules. Yet the final expressions for gravity amplitudes are surprisingly simple and exhibit interesting properties some of which are yet-to-be linked to underlying theoretical or geometric structure. There are multiple modern tools available which make the calculation of graviton amplitudes simpler and manifest important properties of the final result. 

The first of them is the color-kinematics duality. This is motivated by the KLT relations between open and closed string amplitudes \cite{Kawai:1985xq} which extend to Yang-Mills and gravity amplitudes in the low energy limit,
\begin{equation}
{\cal A}^{\rm GR}_n = \sum_{\rho,\sigma} K_{\rho,\sigma} {\cal A}_n^{\rm YM}(\sigma) {\cal A}_n^{\rm YM}(\rho)\,, \label{KLT}
\end{equation}
where we sum over two sets of permutations $\rho$, $\sigma$ of external states of two color-ordered Yang-Mills amplitudes ${\cal A}^{\rm YM}_n$, and the KLT kernel $K_{\rho,\sigma}$ is a certain polynomial in Mandelstam variables $s_{ij}$. It was shown in \cite{Bern:2008qj} that there exists a particular representation of Yang-Mills amplitudes which `squares' into gravity amplitudes at the level of cubic graphs, known as the Bern-Carrasco-Johansson (BCJ) form. The construction also extends to loop amplitudes \cite{Bern:2008qj, Bern:2010ue, Bern:2019prr,Chi:2021mio}. The double copy structure is also manifest in the worldsheet formalism via ambitwistor strings \cite{Geyer:2014fka} and the Cachazo-Ye-Huan (CHY) formula \cite{Cachazo:2013hca}, where the amplitude is expressed as an integral over worldsheet parameters constrained by scattering equations.

The second approach focuses on helicity amplitudes in four dimensions and uses the Britto-Cachazo-Feng-Witten (BCFW) recursion relations \cite{Britto:2004ap, Britto:2005fq} to construct higher-point amplitudes from lower-point ones. The crucial ingredient is the behavior under a large BCFW shift: $\widehat{\widetilde{\lambda}}_n = \widetilde{\lambda}_n + z \widetilde{\lambda}_1,\,\,\widehat{\lambda}_1 = \lambda_1 - z \lambda_n$, 
of the BCFW-shifted amplitude ${\cal A}_n^{\rm GR}(z)$ that vanishes as
\begin{equation}
{\cal A}^{\rm GR}_n(z) = {\cal O}\left(\frac{1}{z^2}\right) \quad \mbox{for} \quad z\rightarrow \infty\,.
\end{equation}
The improved behavior at infinity \cite{Arkani-Hamed:2008bsc} (which is stronger than the $1/z$ scaling of Yang-Mills amplitudes) leads to various BCFW formulae \cite{Cachazo:2005ca,Bedford:2005yy} and many equivalent yet different-looking expressions. The Cauchy formula for the BCFW-shifted amplitude
\begin{equation}
\oint \frac{dz}{z} (1 + \alpha z)A^{\rm GR}_n(z) = 0
\end{equation}
is valid for any $\alpha$ and we can choose $\alpha$ to our liking (unlike in Yang-Mills where we have to set $\alpha=0$ to prevent poles at infinity) or use this freedom to remove one term in the expansion. Interestingly, multi-line shifts do not seem to work in gravity \cite{Bianchi:2008pu} though it is still an open problem \cite{Penante:2012wd,Edison:2019ovj}. The elementary building blocks in any recursion are the two three-point helicity amplitudes, 
\begin{equation}
  \includegraphics[scale=0.5]{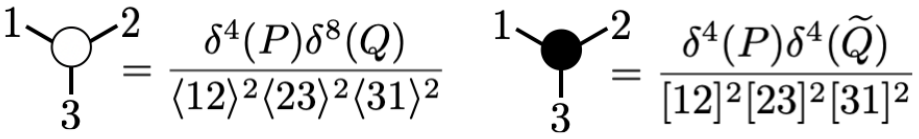}
\end{equation}
which satisfy a simple doubling relation (after stripping off the super-momentum delta function),
\begin{equation}
{\cal A}^{\rm GR}_3(1,2,3) = ({\cal A}^{\rm YM}_3(1,2,3))^2\,.
\end{equation}
At higher-point, it is possible to solve BCFW recursion in a way that uses only ordered objects with a direct connection to Yang-Mills amplitudes. In particular, we can decompose the fully permutation-invariant gravity  amplitude ${\cal A}^{\rm GR}_n$ into a sum of $(n{-}2)!$ ordered amplitudes,
\begin{equation}
{\cal A}^{\rm GR}_n = \sum_{{\cal P}(2,{\dots},n{-}1)} A^{\rm GR}_n(1,2,{\dots},n) \,,
 \label{grav0}
\end{equation}
where labels $n$ and $1$ are special due to the choice of an $(n1)$ shift. These ordered amplitudes can be expressed in terms of dressed planar on-shell diagrams which evaluate to the square of Yang-Mills superfunctions $R_k$ up to a scalar kinematic prefactor $G_k$, 
\begin{equation}
{\cal A}^{\rm GR}_n = \sum_k G_k(s_{ij}) R_k^2\quad\mbox{where}\quad {\cal A}^{\rm YM}_n = \sum_k R_k\,. \label{BCFW1}
\end{equation}
The first expression of this type was found by Elvang and Freedman for MHV amplitudes \cite{Elvang:2007sg} inspired both by KLT and BCFW methods. This later served as motivation for an explicit solution of BCFW for the general N$^k$MHV amplitude in \cite{Drummond:2009ge}. More recently, the recursion was also organized in a new way which makes a direct reference to the cubic graph double copy \cite{Bourjaily:2023tcc}. There are also other closed formulas for MHV amplitudes: the Berends-Giele-Kuijf formula from 1988 \cite{Berends:1988zp}, later shown to be equivalent to the Mason-Skinner formula \cite{Mason:2009afn}, the inverse-soft factor construction \cite{Bern:1998sv, Nguyen:2009jk}, or for more general amplitudes the embedding of BCFW recursion into the language of gravity on-shell diagrams \cite{Arkani-Hamed:2012zlh, Herrmann:2016qea,Heslop:2016plj,Farrow:2017eol,Armstrong:2020ljm}. At loop-level, poles at infinity are generally present, though there are many unexpected cancellations and improved large-momentum scalings \cite{Arkani-Hamed:2008owk,Edison:2019ovj,Bourjaily:2018omh,Herrmann:2018dja,Brown:2022wqr}. There are also very interesting twistor-string inspired \cite{Cachazo:2012da,Cachazo:2012pz,Cachazo:2012kg,Adamo:2012xe,Cachazo:2013zc,Skinner:2013xp} and matrix-representation \cite{Cheung:2012jz,He:2012er,Feng:2012sy} approaches to gravity amplitudes.

Note that the KLT formula (\ref{KLT}) is different from the double-copy-inspired BCFW expressions (\ref{BCFW1}) -- in the former we work with products of Yang-Mills amplitudes with different orderings, while in the latter we square Yang-Mills building blocks with fixed ordering (and sum over permutations). Yet they share the similar idea of building gravity amplitudes (with permutational symmetry) from Yang-Mills amplitudes (with cyclic symmetry). 

There is a different approach to the problem, that is much less developed at the moment, which attempts to exhibit the permutational symmetry of the amplitude while necessarily loosing the manifest double copy connection. The most interesting result on this front is the Hodges formula \cite{Hodges:2012ym} for $n$-point MHV ($k{=}0$) amplitudes, which takes a strikingly simple form
\begin{align}
    \mathcal{A}_{n,0}^{\rm GR} = \frac{|\Phi^{abc}|}{\ang{ab}^2\ang{bc}^2 \ang{ca}^2}\,, \label{Hodges}
\end{align}
where $|\Phi^{abc}|$ is the determinant of $\Phi$ with rows and columns $a$, $b$, $c$ removed. The matrix $\Phi$ has components
\begin{align}
    \Phi_{ij} = \frac{\sq{ij}}{\ang{ij}}, && \Phi_{ii} = \sum_{j\ne i} \frac{[ij]\ang{jk}\ang{jl}}{\ang{ij}\ang{ik}\ang{il}}\,, 
\end{align}
where $\lambda_k$ and $\lambda_l$ are reference spinors and $\Phi_{ii}$ is the soft factor for particle $i$. The formula is independent of the choice of rows and reference spinors. This fascinating expression is manifestly permutation-invariant before fixing the reference rows and columns $a,b,c$ to be removed (there is also a generalized version with six indices and no apparent double poles), but the explicit kinematic formulae loose this manifest invariance due to the various ways momentum conservation may be implemented. This is a consequence of the basic fact that momentum conservation cannot be imposed democratically and one of the momenta can be eliminated from a kinematic expression. In the simple case of the five-point MHV amplitude
\begin{equation}
{\cal A}_{5,0}^{\rm GR} = \frac{N_5\,\delta^4(P)\delta^8(Q)}{\la12\ra\la13\ra\la14\ra\la15\ra\la23\ra\la24\ra\la25\ra\la34\ra\la35\ra\la45\ra}
\end{equation}
the numerator $N_5\,{=}\,\la 12\ra[23]\la34\ra[41]{-}[12]\la23\ra[34]\la41\ra$ is equal to ${\rm Tr}(p_1p_2p_3p_4)$, which is totally antisymmetric in all five labels. There is no form of $N_5$ that manifests all symmetries -- one momentum must be chosen and eliminated. Additionally, Hodges formula holds only for MHV configurations. One of the authors of this letter found an expression for NMHV helicity amplitude ${\cal A}_{n,1}^{\rm GR}(1^-2^-3^-4^+{\dots}n^+)$ that manifests (the full) $S_3\times S_{n{-}3}$ symmetry of this amplitude, but there was no obvious generalization to the supersymmetric case \cite{Trnka:2020dxl}.

In this letter, we will present a new method to express N$^k$MHV amplitudes in terms of simple building blocks, (dressed) one-loop leading singularities. This is motivated by the Britto-Cachazo-Feng (BCF) recursion \cite{Britto:2004ap} for gluon amplitudes, now using additional properties of gravity amplitudes: bonus relations (a consequence of the $1/z^2$ behavior under large BCFW shifts) and also a new type of relation coming from certain zeroes of the amplitude in the collinear region. In the end, we express the amplitude in terms of new building blocks, very reminiscent of the expansion of gluon amplitudes in terms of Yangian invariants, but never introduce any ordering of the external states -- our objects manifest their own permutational symmetry (as a subset of the total permutational symmetry of the amplitude). We focus on the NMHV case, but also illustrate the generalization to higher $k$. In the end, we discuss a curious connection between these new objects and Grassmannian geometry, motivated by such a connection between Yangian invariants (as building blocks for gluon amplitudes) and dlog forms on the cells in the positive Grassmannian.

\vspace{-0.3cm}

\section{Gluon Amplitudes From Triple Cuts}

\vspace{-0.3cm}

The BCFW-shifted color-ordered amplitude ${\cal A}_{n}^{\rm YM}(z)$ can be interpreted as a one-loop triple cut,
\begin{equation}
  \includegraphics[scale=0.52]{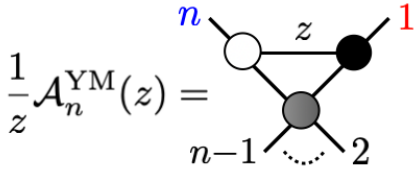}\label{grt0}
\end{equation}
where we added a BCFW bridge \cite{Arkani-Hamed:2012zlh} to the grey blob representing the unshifted amplitude ${\cal A}_n^{\rm YM}$. In these figures all legs are on-shell, both internal (cut propagators) and external. The momentum flow in the bridge (horizontal internal line) is linear in the shift parameter, $P=z\lambda_1\widetilde{\lambda}_n$. The residue theorem for the triple cut function
\begin{equation}
  \includegraphics[scale=0.32]{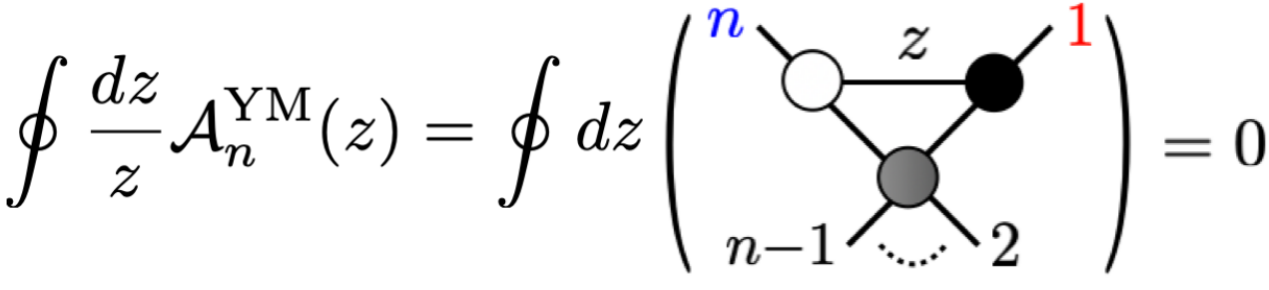}\label{GRT}
\end{equation}
can be interpreted diagramatically as a relation between on-shell functions. We can use (\ref{GRT}) to express the original unshifted amplitude ${\cal A}_n^{\rm YM}$ (residue at $z=0$) in terms of other residues which originate from factorization channels of ${\cal A}_n^{\rm YM}(z)$. For the $k=1$ i.e. NMHV case, 
\begin{equation}
  \includegraphics[scale=0.27]{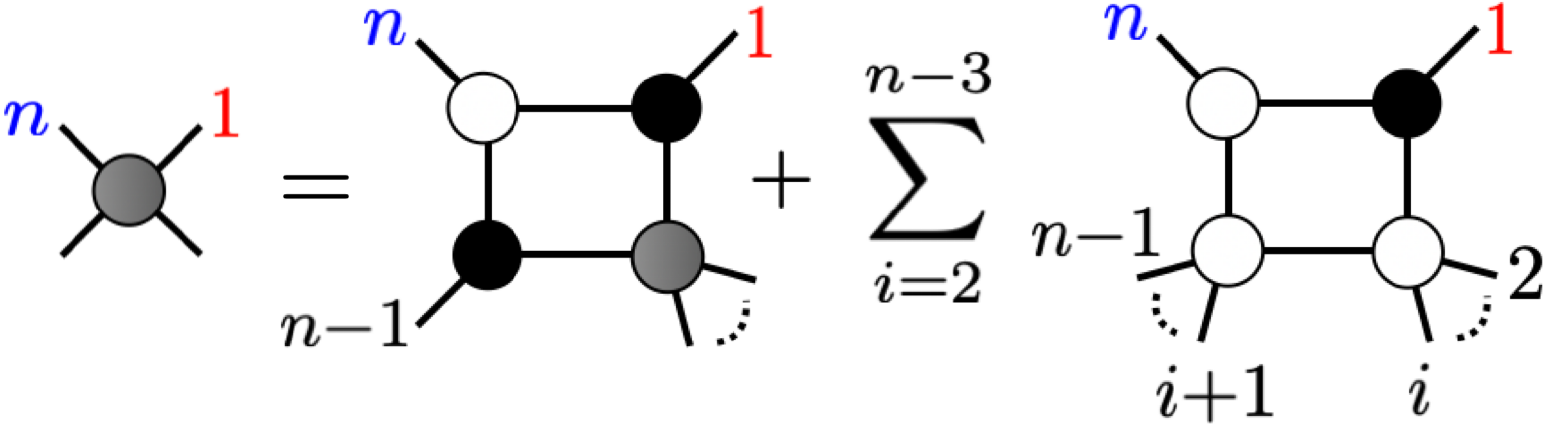}\label{grt1}
\end{equation}
where the dark gray blob is the NMHV amplitude ${\cal A}_{n,1}^{\rm YM}$ and the terms of the RHS are one-loop leading singularities \cite{Britto:2004ap}. In the next recursive step, we could express the gray blob in the second term by the same sum (\ref{grt1}) but now with $n{-}1$ external legs. As a result, we would get higher-loop leading singularities in the expansion for ${\cal A}_{n,1}^{\rm YM}$. There is also a way to rewrite the formula for ${\cal A}_{n,1}^{\rm YM}$ from (\ref{grt1}) only in terms of one-loop leading singularities. We remove the second term in (\ref{grt1}) by using a global residue theorem on the following triple cut function,
\begin{equation}
  \includegraphics[scale=0.45]{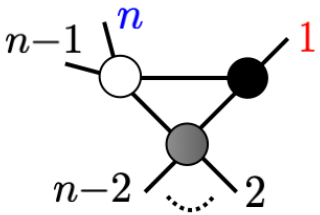}
\end{equation}
which produces a relation between leading singularities,
\begin{equation}
 \hspace{-0.3cm}\raisebox{-.83cm}{\includegraphics[scale=0.35]{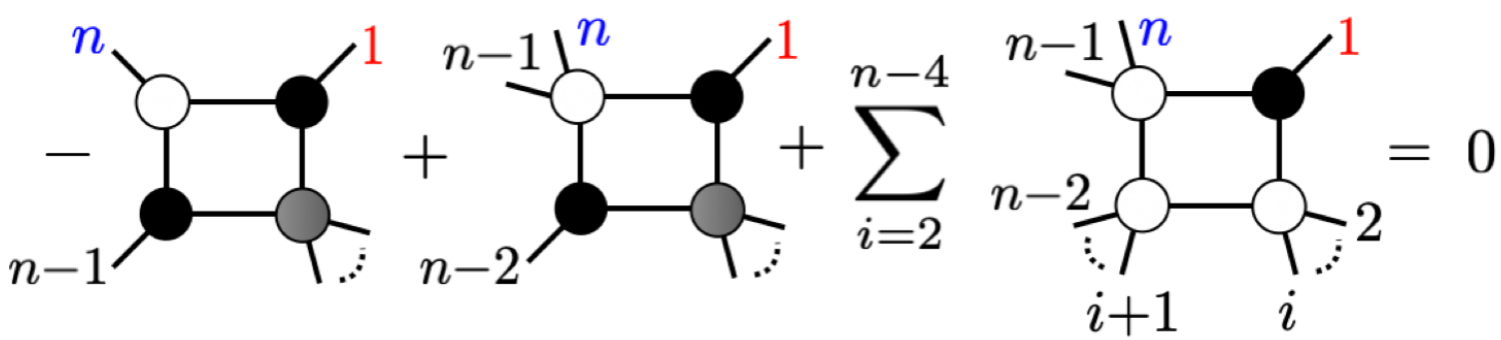}} \label{grt3}
\end{equation}
We continue this procedure and remove the second term in (\ref{grt3}) using another residue theorem, where legs $n{-}2$, $n{-}1$ and ${\color{blue} n}$ are now attached to the same vertex. After all terms with lower-point NMHV blobs are removed, we get the final result for the $n$-point NMHV amplitude
\begin{equation}
\includegraphics[scale=0.37]{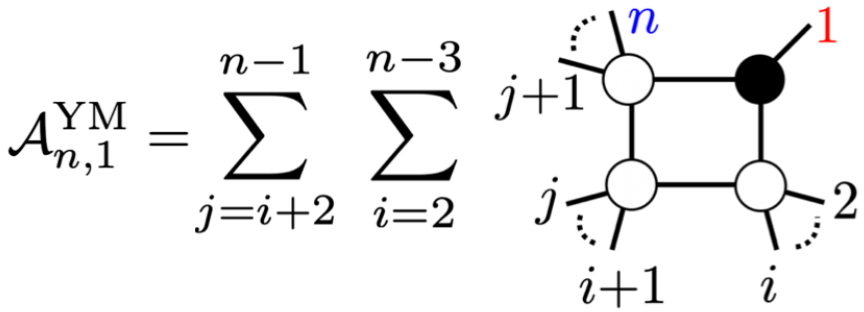}\label{YMform}
\end{equation}
where each term in the sum is famously an $R$-invariant ${\cal R}_{1,i{+}1,j{+}1}$ \cite{Drummond:2008vq}. For higher $k$, we can write similar formulae which reduce terms in the recursive expansion to $k$-loop leading singularities with only MHV (white) vertices and $k$ three-point $\overline{\rm MHV}$ (black) vertices.
Each term corresponds to a canonical dlog form on a cell in the positive Grassmannian $G_+(k,n)$ \cite{Arkani-Hamed:2012zlh}. 

\vspace{-0.3cm}

\section{New Gravity Formulae}

\vspace{-0.3cm}

The BCFW bridge in the triple cut function for ${\cal A}_{n}^{\rm GR}(z)$ now has an additional factor of $s_{n1}$,
\begin{equation}
\includegraphics[scale=0.5]{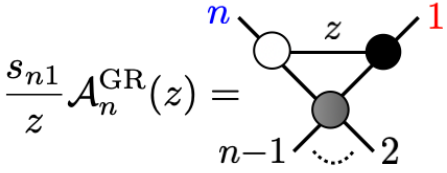}\label{GRres}
\end{equation}
The residue theorem (\ref{GRT}) for $k{=}1$ (NMHV) leads to 
\begin{equation}
\hspace{-0.1cm}  \includegraphics[scale=0.48]{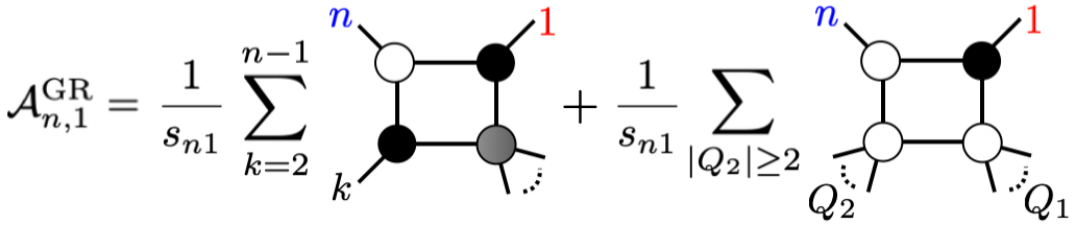}\label{grav}
\end{equation}
where the corner $Q_2$ has at least two external legs. We would like to follow the same procedure as in the Yang-Mills case and get rid off the first sum on the right hand side of (\ref{grav}) which contains a lower-point NMHV amplitude. Because of the scaling ${\cal A}_n^{\rm GR}(z) \sim 1/z^2$ at $z\rightarrow\infty$, we can use the residue theorem for the gravity triple cut function (\ref{GRT}) and also the same relation multiplied by $z$ -- these are also called `bonus relations' \cite{Arkani-Hamed:2008bsc,He:2010ab,Spradlin:2008bu}, written using the triple cut as
\begin{equation}
  \includegraphics[scale=0.32]{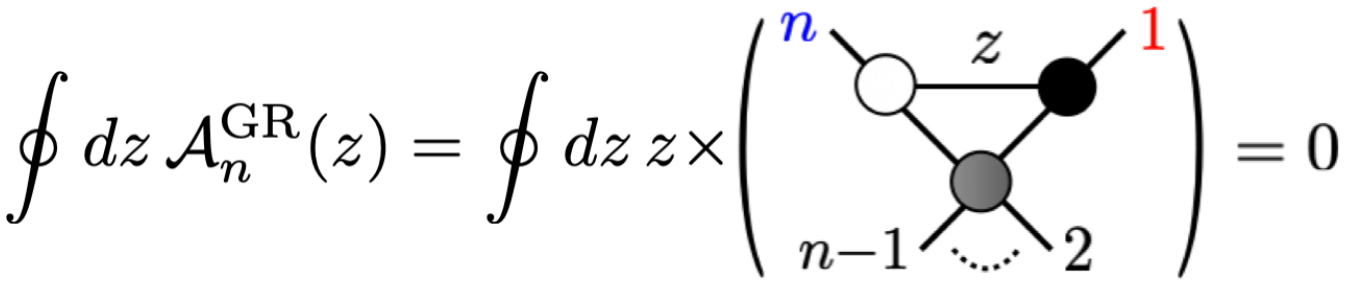}\label{eq:bonus}
\end{equation}
which relates different one-loop leading singularities in (\ref{grav}). The residue on $z{=}0$ (which is ${\cal A}_n^{\rm GR}$) is absent in \eqref{eq:bonus}. We now consider another set of triple cut functions,
\begin{equation}
  \includegraphics[scale=0.53]{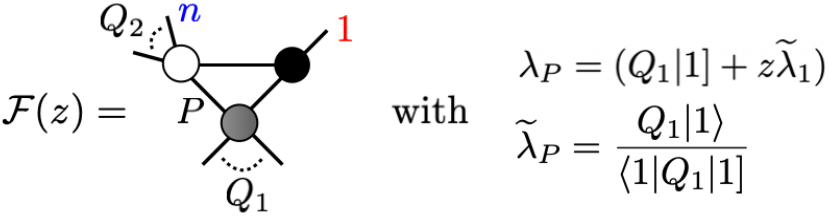}\label{grt5}
\end{equation}
It was conjectured in \cite{Herrmann:2016qea} that any one-loop triple cut function scales as ${\cal F}(z)\sim 1/z^3$ for $z\rightarrow\infty$ (analogue of bonus relations). As a result, at least in principle, we can use residue theorems such as
\begin{equation}
\oint dz\,(1{+}\alpha z){\cal F}(z) = 0
\end{equation}
to further relate terms in (\ref{grav}) with the constraint that we do not introduce any new leading singularities with NMHV vertices in order to preserve the form of the answer, similar to the Yang-Mills case (\ref{YMform}). 

As it turns out, this is not enough and it is \emph{not} possible to get rid off all terms in the first sum in (\ref{grav}). In the Yang-Mills case (\ref{grt1}), there was only one such term but now we have $n{-}2$ of them, and we do not have enough relations to eliminate all the terms even if we use the improved $1/z^2$ behavior. With no further relations available, the analogue of (\ref{YMform}) would not exist and we would only get a generic expansion in terms of higher-loop leading singularities (or on-shell diagrams when rewriting everything in terms of three-point on-shell vertices).

However, there is extra information about the \emph{zeroes} of triple cuts (\ref{grt5}) that we can use to write additional relations. This comes from the collinear behavior of the amplitude, also known as the splitting function \cite{Bern:1998sv}, applied to on-shell diagrams and leading singularities in \cite{Herrmann:2016qea}. In particular, the function ${\cal F}(z)$ in (\ref{grt5}) \emph{vanishes} for $z=0$ for all triple cuts if $Q_2$ has at least two external legs (one of them being ${\color{blue} n}$). Hence, we get a more general relation
\begin{equation}
\oint dz \frac{(1+\alpha z)}{z} {\cal F}(z) = 0 \label{bonus2}
\end{equation}
which gets no contribution from infinity. Because of the multiplicative \emph{and} divisive factors in the integrand, we refer to relations between leading singularities from (\ref{bonus2}) as \emph{double bonus relations}. This does not hold if $Q_2$ has only one leg ${\color{blue} n}$, i.e. for the actual shifted amplitude (\ref{GRres}). In this case the triple cut function has a pole (rather than a zero) as also evident from (\ref{GRres}). Note that there are further relations that stem from
\begin{equation}
\oint dz \frac{(1+\alpha z + \beta z^2)}{z} {\cal F}(z) = 0\,,
\end{equation}
but we do not study their consequences because \eqref{bonus2} will be sufficient for our purposes. Indeed using (\ref{grt5}) and (\ref{bonus2}) for fixed $\alpha$ we can rewrite the first sum in ${\cal A}_{n,1}^{\rm GR}$ as
\begin{equation}
  \includegraphics[scale=0.54]{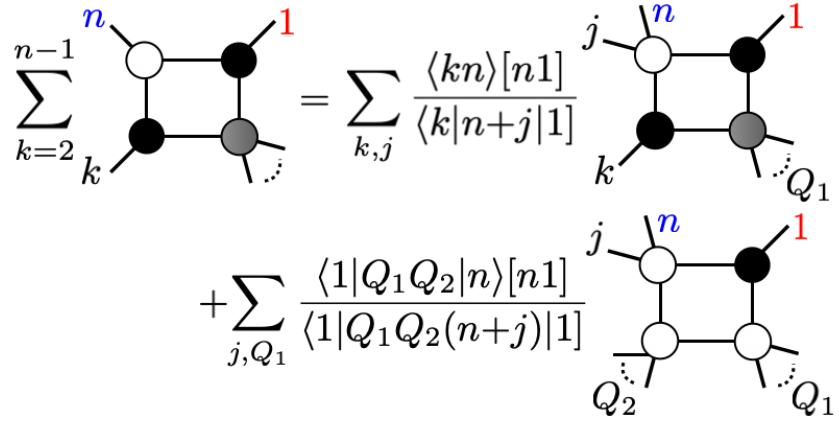}\label{new2}
\end{equation}
Next, we rewrite the first sum on the right hand side of (\ref{new2}) using the same type of residue theorem, now with two legs $\{n,j\}$ in the corner replaced by three legs $\{n,j,i\}$, and so on. Exploiting all these relations we get
\begin{equation}
  \includegraphics[scale=0.44]{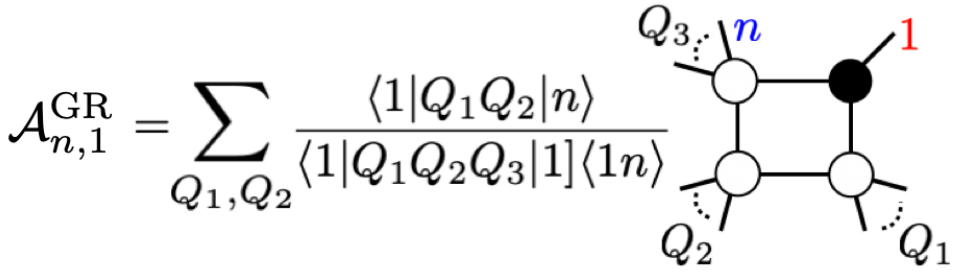}\label{new3}
\end{equation}
where ${\cal J}=\langle 1|Q_1Q_2Q_3|1]$ is a compact way to write the Jacobian of this generic one-loop triple cut. Note that the blob $Q_2$ needs to have at least two external legs. This is a precise analogue of the Yang-Mills expression in terms of $R$-invariants (\ref{YMform}) as far as the structure is concerned. The difference is that the one-loop leading singularities are dressed with kinematic prefactors. Note that the leg {{\color{blue} $n$} is fixed to be in the blob $Q_3$ -- in the ordered case of (\ref{YMform}) this was automatic. Momentum conservation requires $Q_1+Q_2+Q_3+p_1 = 0$. The formulae simplify slightly if we rewrite the prefactor using the solutions to on-shell internal momenta $P_1$, $P_2$, $P_3$ and $P_4$,
\begin{equation}
  \includegraphics[scale=0.55]{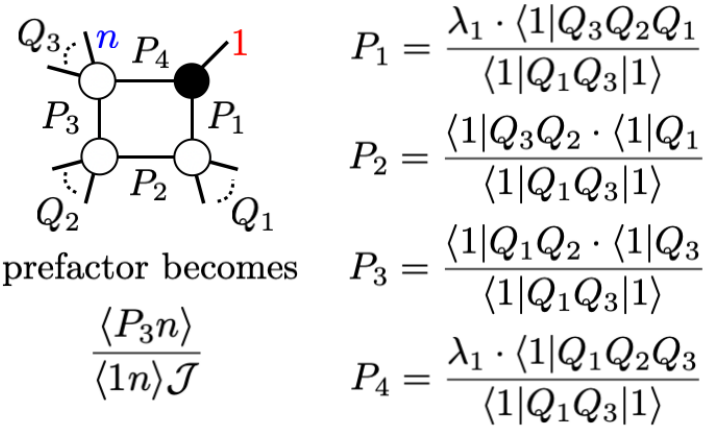}\label{new4}
\end{equation}
where we set $\lambda_{P_1}=\lambda_{P_4}=\lambda_1$, $\lambda_{P_2}=\la 1|Q_3Q_2$ and $\lambda_{P_3}=\la 1|Q_1Q_2$. After solving for internal on-shell momenta the (un-dressed) one-loop leading singularity evaluates to
\begin{equation}
\int d^8\widetilde{\eta}_{P_1}d^8\widetilde{\eta}_{P_2}
d^8\widetilde{\eta}_{P_3}d^8\widetilde{\eta}_{P_4}\, \frac{A_1A_2A_3A_4}{\cal J}\,,
\end{equation}
where we denoted amplitudes in individual corners
\begin{align*}
&A_1={\cal A}^{\rm GR}_{\rm MHV}(P_1,\{Q_1\},P_2), \  A_2={\cal A}^{\rm GR}_{\rm MHV}(P_2,\{Q_2\},P_3)\,,\\
&A_3={\cal A}^{\rm GR}_{\rm MHV}(P_3,\{Q_3\},P_4),\  A_4={\cal A}_{\overline{\rm MHV}}^{\rm GR}(P_4,{\color{red} 1},P_1)\,.
\end{align*}
We integrate over the fermionic variables $\widetilde{\eta}_{P_j}$ and get
\begin{equation}
    \delta^4(P)\delta^{16}(Q)\delta^8(\Xi) \times \widetilde{A}_1\widetilde{A}_2\widetilde{A}_3 \times {\cal J} \times \la  1|Q_1Q_3|1\ra^6 \label{Ginv}
\end{equation}
where $\widetilde{A}_k$ are the bosonic parts of three MHV amplitudes -- we can for example use the Hodges formula (\ref{Hodges}) with $\lambda_{P_k}$, $\widetilde{\lambda}_{P_k}$ from (\ref{new4}) where the last factor in (\ref{Ginv}) is the normalization of $P_k$. It is easy to write an explicit closed form for each one-loop leading singularity using Hodges determinants, and hence the whole amplitude ${\cal A}_{n,1}^{\rm GR}$ for any number of points. The delta function $\delta^8(\Xi)$ is the same as in the Yang-Mills case (up to ${\cal N}=4 \rightarrow 8$) and equal to
\begin{equation}
\Xi = \sum_{j\in Q_1} Q_2^2\la 1j\ra \widetilde{\eta}_j + \sum_{k\in Q_2}\la k|Q_2Q_1|1\ra \widetilde{\eta}_k \,.   
\end{equation}
The same procedure applies to higher $k$. For N$^2$MHV, we can use the general residue theorem of the form (\ref{bonus2}) on a complete set of N$^2$MHV triple cuts with one 3-point $\overline{\rm MHV}$ attached to the external leg {\color{red} 1} to eliminate terms with a lower point N$^2$MHV vertex and write the BCFW formula as
\begin{align}
\input{fig21}    
\end{align}
where as before the gray vertex represents the NMHV amplitude, now given by the formula \eqref{new3}. This is an analogue of the N$^2$MHV Yang-Mills formula \cite{Drummond:2008cr} in terms of two building blocks $A$ and $B_1$+$B_2$ (which correspond to two different types of $k=2$ Yangian invariants) as we also have in (\ref{new4a}). We can further express the gray NMHV vertices in terms of (anti-)MHV vertices, and get
\begin{equation}
  \includegraphics[scale=0.42]{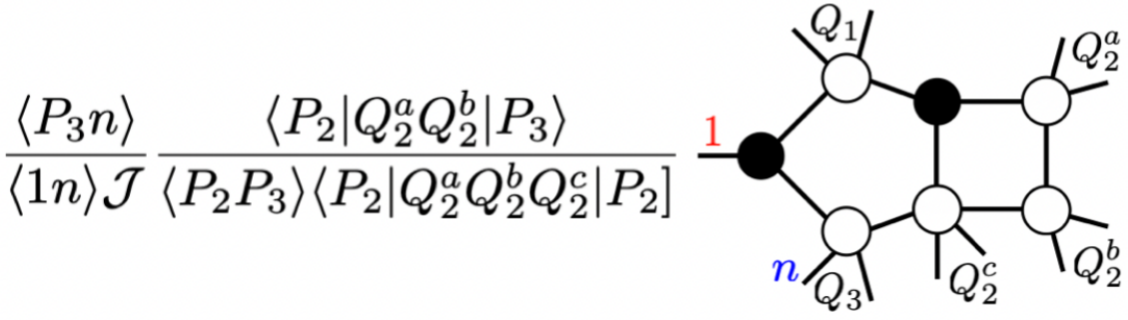}\label{new5}
\end{equation}
for the first term, where the NMHV prefactor in \eqref{new3} is evaluated via the replacements $1\to P_2$, $n\to P_3$, $Q_1\to Q_2^a$, $Q_2\to Q_2^b$ and $Q_3\to Q_2^c$. Similarly, the second contribution to the N$^2$MHV amplitude gives,
\begin{equation}
  \includegraphics[scale=0.42]{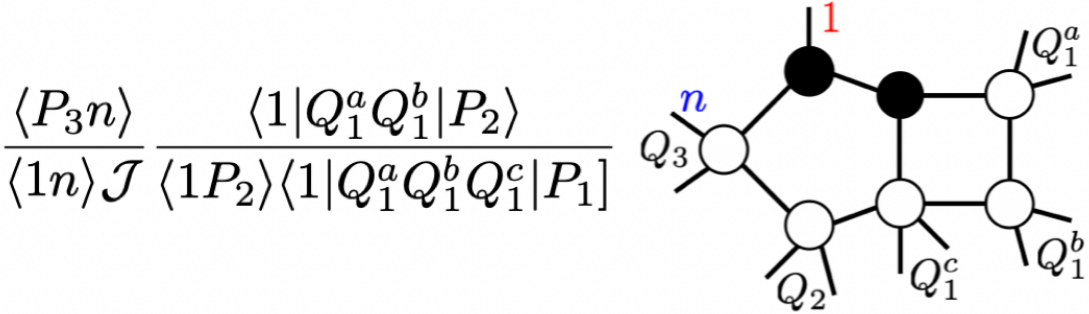}\label{new6}
\end{equation}
Again the prefactor in \eqref{new3} is evaluated via replacements $1\to P_1$, $n\to P_2$, $Q_1\to Q_1^a$, $Q_2\to Q_1^b$ and $Q_3\to Q_1^c$. The momenta $P_i$ refer to those defined in (\ref{new4}). We can perform a similar calculation for N$^k$MHV amplitudes, which will use expressions for the N$^{k{-}1}$MHV amplitude in terms of (anti-)MHV vertices,
\input{fig25}
Using the same procedure as before, we can express the gray blobs in terms of minimal diagrams with only three-point black and higher-point white vertices. As a result, the N$^k$MHV gravity amplitude can be then expressed in terms of decorated (with kinematical prefactors) $k$-loop leading singularities with $k$ black vertices and $2k{+}1$ white vertices. In the analogous Yang-Mills case each term represents one higher-$k$ Yangian invariant.

Finally, let us emphasize that the NMHV formula (\ref{new3}) and the higher-$k$ generalizations do not contain a sum over permutations of ordered expressions. Rather each term in (\ref{new3}) is built from permutation-invariant objects (lower-point MHV amplitudes), though the complete $S_n$ permutation symmetry is only carried by the whole sum.

\vspace{-0.3cm}

\section{Towards Grassmannian Geometry}

\vspace{-0.3cm}

In ${\cal N}=4$ SYM theory, BCFW terms are the same objects as one-loop leading singularities. In the dual Grassmannian formulation, they can be obtained as dlog forms on the cells in the positive Grassmannian $G_+(k{+}2,n)$. Furthermore, the whole sum (\ref{YMform}) can then be interpreted geometrically as a triangulation of an underlying Amplituhedron geometry \cite{Arkani-Hamed:2013jha,Arkani-Hamed:2017vfh,Damgaard:2019ztj,Ferro:2022abq}. In the particular case of $n$-point NMHV, each term in (\ref{YMform}) corresponds to a cell in the positive Grassmannian $G_+(3,n)$ which can be represented by a (convex) configuration of $n$ ordered points in $\mathbb{P}^2$, localized on three lines,
\begin{equation}
  \includegraphics[scale=0.52]{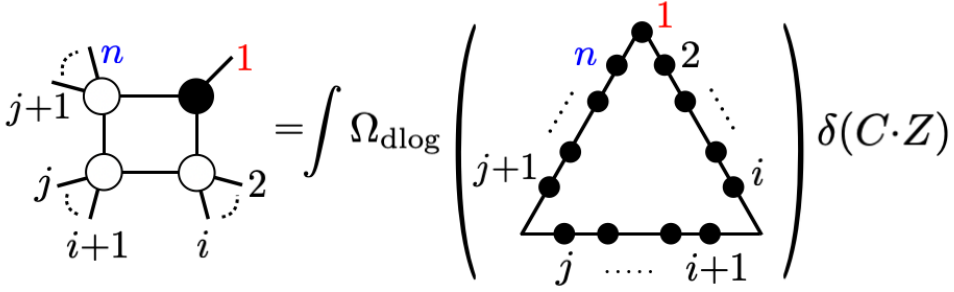}\label{geom1}
\end{equation}
where $\Omega_{\rm dlog}$ is a dlog form in the parameters of the Grassmannian matrix $C$, and $\delta(C\cdot Z)$ denotes a collection of delta functions that solve for the Grassmannian parameters in terms of kinematic data $\lambda$, $\widetilde{\lambda}$, $\widetilde{\eta}$. The dlog form is very specific to the ${\cal N}=4$ SYM theory (but extensions exist for ${\cal N}<4$ SYM theory \cite{Arkani-Hamed:2012zlh}). Our ultimate goal is to find an analogous formula for gravity amplitudes: $\Omega_{\rm dlog}$ must be replaced by some other $\Omega_{\rm GR}$ which reflects the singularity structure of gravity amplitudes, and the points in $\mathbb{P}^2$ should not exhibit any ordering (see \cite{Paranjape:2022ymg} for study of such configurations for non-adjacent BCFW shifts in $\mathcal{N}=4$ Yang-Mills theory). 

In the gluon case, the first non-trivial example is a convex configuration of six points in a plane where points $1,2,3$ are on a line (modulo cyclic permutations).
\begin{equation}
  \includegraphics[scale=0.37]{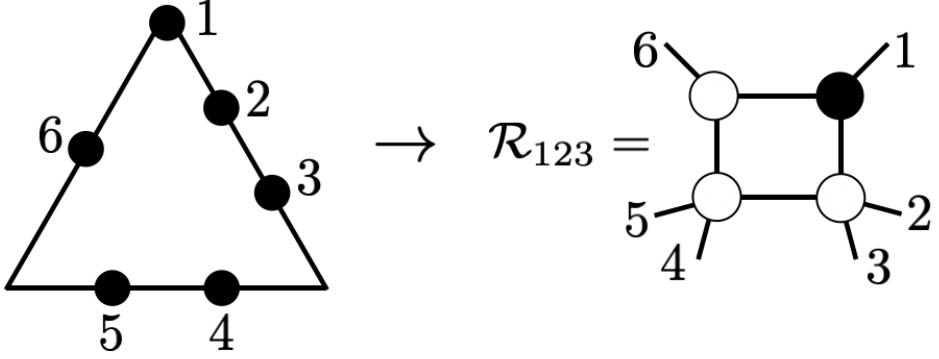}\label{geom2}
\end{equation}
Here we denoted ${\cal R}_{123}$ one of the $R$-invariants, the building blocks of six-point NMHV amplitudes. There are three equivalent ways to draw this configuration (corresponding to a rotation of lines) which relates three one-loop leading singularities, 
\begin{equation}
  \includegraphics[scale=0.33]{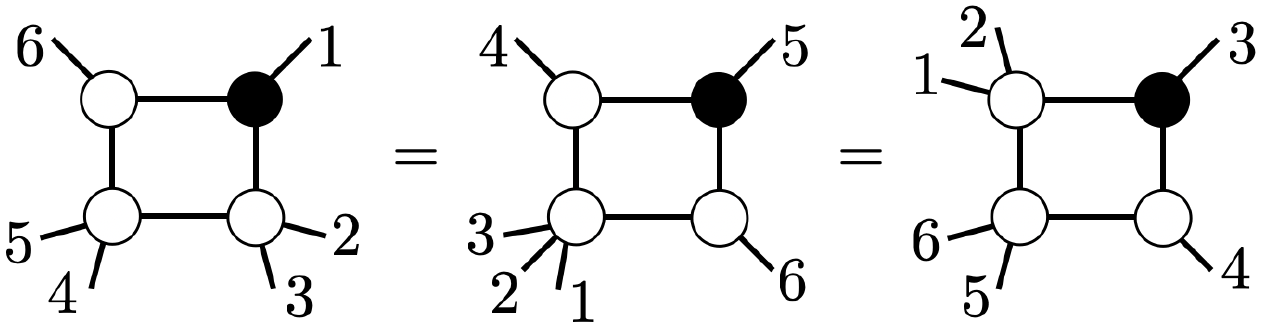}\label{geom3}
\end{equation}
An analogous object can be defined for gravity, now by summing all one-loop leading singularities which (in analogy with Yang-Mills) could correspond to six (unordered) points in $\mathbb{P}^2$ with points $1,2,3$ on a line. There are two types of leading singularities that correspond to such configurations. First, we get the following collection,
\begin{equation}
  \includegraphics[scale=0.35]{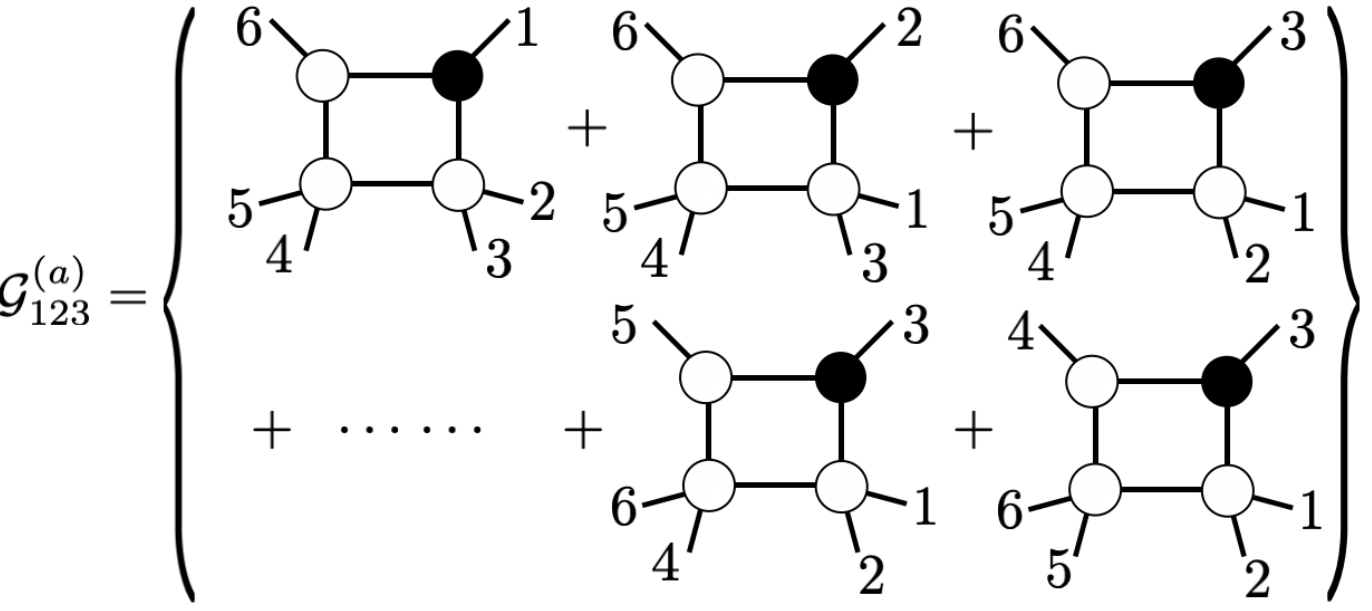}\label{geom4}
\end{equation}
Using (\ref{Ginv}) we can evaluate it to
\begin{equation}
{\cal G}_{123}^{(a)} = \sum_{S_{123}\times S_{456}}
\frac{[23]\la45\ra s_{61}}{\begin{array}{c}s_{123}\la12\ra\la23\ra\la13\ra[45][56][46]\\ \la1|23|4]\la1|23|5]\la2|13|6]\la3|12|6]\end{array}}
\end{equation}
where we omitted in the numerator the delta functions $\delta^4(P)\delta^{16}(Q)\delta^8([45]\widetilde{\eta}_6 {+} [56]\widetilde{\eta}_4 {+} [64]\widetilde{\eta}_5)$ and used a short notation $\la a|bc|d]$ for $ \la ab\ra[bd] {+} \la ac\ra[cd]$. There is another way to draw the configurations of points, that gives us the following collection of leading singularities,
\begin{equation}
  \includegraphics[scale=0.37]{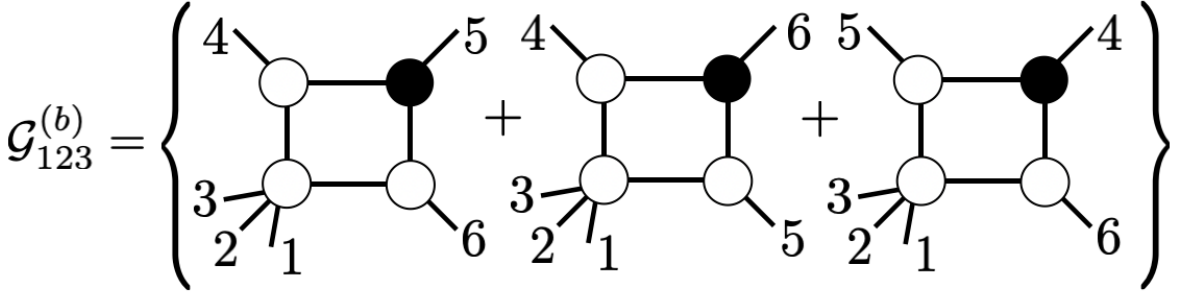}\label{geom5}
\end{equation}
which is a sum of three terms (each of which manifest permutation symmetry in $1,2,3$),
\small
\begin{equation}
{\cal G}_{123}^{(b)} {=} \sum_{S_{456}} \frac{\la45\ra\la56\ra(\la12\ra[23]\la3|45|6][14]{-}[12]\la23\ra[34]\la1|45|6])}{\begin{array}{c}s_{123}\la12\ra\la23\ra\la13\ra[45][56]\la1|23|4]\la2|13|4] \\ \la3|12|4]\la1|23|6]\la2|13|6]\la3|12|6]\end{array}}
\end{equation}
\normalsize
with the same set of delta functions. Explicit check shows
\begin{equation}
    {\cal G}_{123}^{(a)} = {\cal G}_{123}^{(b)} \equiv {\cal G}_{123}
\end{equation}
which is an analogue of (\ref{geom3}) in Yang-Mills. This relation can be proven by residue theorems for certain triple cuts (with only MHV vertices). In the Yang-Mills case, the $R$-invariants also satisfy an important six-term identity,
\begin{equation}
    {\cal R}_{123} + {\cal R}_{234} + {\cal R}_{345} + {\cal R}_{456} + {\cal R}_{561} + {\cal R}_{612} = 0 \label{Rrel}
\end{equation} 
where ${\cal R}_{ijk}$ corresponds to a configuration of six points where points $i,j,k$ are now on the same line (for example ${\cal R}_{123} \equiv {\cal R}_{1,3,6}$ in the usual ${\cal R}$-invariant notation). This is a very nice consequence of the residue theorem in the Grassmannian representation \cite{Arkani-Hamed:2009ljj,Arkani-Hamed:2009kmp,Arkani-Hamed:2009pfk,Arkani-Hamed:2012zlh}. The new objects ${\cal G}_{abc}$ satisfy an analogous formula,
\begin{equation}
    \sum_{S_6} {\cal G}_{abc} = 0 \label{Grel}
\end{equation}
which is a 20-term identity for ${\cal G}_{abc}$ (and hence a 60-term identity for one-loop leading singularities using (\ref{geom5})). This bears a striking similarity to (\ref{Rrel}) with $R$-invariants and further is very suggestive of the existence of a Grassmannian construction, though the actual building blocks for tree-level gravity amplitudes (\ref{new3}) are dressed with kinematical factors (\ref{new4}). Hence these prefactors should play a crucial role in the putative geometric construction, which we explore in upcoming work \cite{inprogress:Oktem}. 

\vspace{-0.3cm}

\section{Conclusion and Outlook}

\vspace{-0.3cm}

In this letter, we present new expressions for $n$-point graviton amplitudes in terms of canonical building blocks. In the NMHV case, these are one-loop leading singularities dressed with kinematical prefactors, very similar to the Yang-Mills case where the same (but un-dressed) objects are term-wise equal to $R$-invariants. The analogy goes even further and we show that for the six-point case the (un-dressed) objects satisfy new relations, similar to a six-term relation between $R$-invariants. 

The crucial question for the future is the role of the kinematical dressing, how it fits into the story of a (putative) Grassmannian geometry, residue theorems and bonus relations, and how to show in some geometric way that different BCFW formulae (after being rewritten in an appropriate form using our double bonus relations) give the same amplitude. One particular direction is the careful study of spurious pole cancellation. In the gluon case \cite{Hodges:2009hk} this lead to important insights and the eventual discovery of the Amplituhedron \cite{Arkani-Hamed:2013jha}. Another path is to find a closer link with the formula for the $n$-point NMHV fixed helicity amplitude presented in \cite{Trnka:2020dxl}, which makes manifest complete permutational symmetry (separately in the labels of positive and negative helicity external states). The interplay between the formula given in \cite{Trnka:2020dxl}, the ordered expressions for gravity \cite{Bourjaily:2023tcc} and our new BCFW expression (\ref{new3}), together with a deeper understanding of the Hodges MHV formula (which is implicitly used in our building blocks as the white MHV vertices) might bring us closer to the discovery of a putative Gravituhedron geometry.

\medskip

{\it Acknowledgements}: We thank Nima Arkani-Hamed, Jacob Bourjaily, Taro Brown, Song He and Umut Oktem for very useful discussions. This work is supported by GA\-\v{C}R 21-26574S, DOE grant No. SC0009999 and the funds of the University of California.

\bibliography{mainbib}
\bibliographystyle{apsrev4-1}

\end{document}

%% file: header_data.tex
\titleformat{\section}{\centering\normalsize\normalfont\bf}{\thesection}{0em}{}
\hypersetup{pdftitle={},pdfcreator={},linkcolor=[rgb]{0.15,0.35,0.75},colorlinks=true,citecolor=[rgb]{0.675,0,0.2},urlcolor=[rgb]{0.15,0.35,0.65}}
\renewcommand{\phi}{\varphi}

\definecolor{nmhv}{rgb}{0.95,0.55,0.55}
\definecolor{mhvblue}{rgb}{0.6,0.6,0.7765}
\definecolor{ampgrey}{rgb}{0.9,0.9,0.9}
\definecolor{hblue}{rgb}{0,0,0.575}
\definecolor{hred}{rgb}{0.575,0.0,0.225}
\definecolor{hgreen}{rgb}{0.0,0.4,0.2}
\definecolor{hteal}{rgb}{0.0,0.545,0.7451}
\definecolor{perm}{rgb}{0.1,0.45,0.85}
\definecolor{unord}{rgb}{0,0,0}
\definecolor{ord}{rgb}{0,0,0.575}
\definecolor{anchorLeg}{rgb}{0.575,0.0,0.225}
\definecolor{fRed}{rgb}{0.48,0.02824,0.18824}

%% file: fig21.tex
{\cal A}_{n,2}^{\rm GR} = \sum_{Q_1,Q_2} \frac{\ang{P_3n}}{\ang{1n} \mathcal{J}}\left(
\begin{tikzpicture}[scale=0.5,baseline={(0,0.375cm)}]
    \draw[line width=1pt] (0,0)--(1.8,0);
    \draw[line width=1pt] (0,0)--(0,1.5);
    \draw[line width=1pt](1.8,0)--(1.8,1.5);
    \draw[line width=1pt](1.8,1.5)--(0,1.5);
    \draw[line width=1pt](0,0)--(-0.25,-0.75);
    \draw[line width=1pt](0,0)--(-0.75,-0.25);
    \draw[line width=1pt](1.8,0)--(2.55,-0.25);
    \draw[line width=1pt](1.8,0)--(2.05,-0.75);
    \draw[line width=1pt](0,1.5)--(-0.25,2.25) node [at end, above, blue] {\small $n$};
    \draw[line width=1pt](0,1.5)--(-0.75,1.75);
    \draw[line width=1pt](1.8,1.5)--(2.5,2) node [at end, above right, red] {\small 1};
    \draw[black,fill=gray] (0,0) circle (2.5ex);
    \draw[black,fill=white] (1.8,0) circle (2.5ex);
    \draw[black,fill=white] (0,1.5) circle (2.5ex);
    \draw[black,fill=black] (1.8,1.5) circle (2.5ex);
    \draw[dotted] (-0.2,-0.6) to [out=-120,in=-160] (-0.6,-0.2);
    \node[below left] at (-0.25,-0.25) {\small $Q_2$};
    \draw[dotted] (2,-0.6) to [out=0,in=-90] (2.4,-0.2);
    \node[below right] at (2.25,-0.25) {\small $Q_1$};
    \draw[dotted] (-0.2,2) to [out=180,in=90] (-0.6,1.7);
    \node[above left] at (-0.25,1.75) {\small $Q_3$};
\end{tikzpicture} \hspace{-0.25cm}+ \hspace{-0.25cm}\begin{tikzpicture}[scale=0.5,baseline={(0,0.375cm)}]
    \draw[line width=1pt] (0,0)--(1.8,0);
    \draw[line width=1pt] (0,0)--(0,1.5);
    \draw[line width=1pt](1.8,0)--(1.8,1.5);
    \draw[line width=1pt](1.8,1.5)--(0,1.5);
    \draw[line width=1pt](0,0)--(-0.25,-0.75);
    \draw[line width=1pt](0,0)--(-0.75,-0.25);
    \draw[line width=1pt](1.8,0)--(2.55,-0.25);
    \draw[line width=1pt](1.8,0)--(2.05,-0.75);
    \draw[line width=1pt](0,1.5)--(-0.25,2.25) node [at end, above, blue] {\small $n$};
    \draw[line width=1pt](0,1.5)--(-0.75,1.75);
    \draw[line width=1pt](1.8,1.5)--(2.5,2) node [at end, above right, red] {\small 1};
    \draw[black,fill=white] (0,0) circle (2.5ex);
    \draw[black,fill=gray] (1.8,0) circle (2.5ex);
    \draw[black,fill=white] (0,1.5) circle (2.5ex);
    \draw[black,fill=black] (1.8,1.5) circle (2.5ex);
    \draw[dotted] (-0.2,-0.6) to [out=-120,in=-160] (-0.6,-0.2);
    \node[below left] at (-0.25,-0.25) {\small $Q_2$};
    \draw[dotted] (2,-0.6) to [out=0,in=-90] (2.4,-0.2);
    \node[below right] at (2.05,-0.25) {\small $Q_1$};
    \draw[dotted] (-0.2,2) to [out=180,in=90] (-0.6,1.7);
    \node[above left] at (-0.25,1.75) {\small $Q_3$};
\end{tikzpicture}\right)\label{new4a}

%% file: fig25.tex
\begin{align}
    &\mathcal{A}_{n,k}^{\rm GR} = \sum_{Q_1,Q_2}\frac{\ang{P_3n}}{\ang{1n}\mathcal{J}}\nonumber\\
    &\left(\begin{tikzpicture}[scale=0.5,baseline={(0,0.375cm)}]
    \draw[line width=1pt] (0,0)--(1.8,0);
    \draw[line width=1pt] (0,0)--(0,1.5);
    \draw[line width=1pt](1.8,0)--(1.8,1.5);
    \draw[line width=1pt](1.8,1.5)--(0,1.5);
    \draw[line width=1pt](0,0)--(-0.25,-0.75);
    \draw[line width=1pt](0,0)--(-0.75,-0.25);
    \draw[line width=1pt](1.8,0)--(2.8,-0.33);
    \draw[line width=1pt](1.8,0)--(2.13,-1);
    \draw[line width=1pt](0,1.5)--(-0.25,2.25) node [at end, above, blue] {\small $n$};
    \draw[line width=1pt](0,1.5)--(-0.75,1.75);
    \draw[line width=1pt](1.8,1.5)--(2.5,2) node [at end, above right, red] {\small 1};
    \draw[black,fill=white] (0,0) circle (2.5ex);
    \draw[black,fill=lightgray] (1.8,0) circle (4.5ex);
    \draw[black,fill=white] (0,1.5) circle (2.5ex);
    \draw[black,fill=black] (1.8,1.5) circle (2.5ex);
    \node[black] at (1.8,0) {\scriptsize $k{-}1$};
    \draw[dotted] (-0.2,-0.6) to [out=-120,in=-160] (-0.6,-0.2);
    \node[below left] at (-0.25,-0.25) {\small $Q_2$};
    \draw[dotted] (2.05,-0.95) to [out=0,in=-60] (2.55,-0.28);
    \node[below right] at (2.3,-0.5) {\small $Q_1$};
    \draw[dotted] (-0.2,2) to [out=120,in=90] (-0.6,1.7);
    \node[above left] at (-0.25,1.75) {\small $Q_3$};
\end{tikzpicture}\hspace{-0.25cm}+\hspace{-0.25cm} \begin{tikzpicture}[scale=0.5,baseline={(0,0.375cm)}]
    \draw[line width=1pt] (0,0)--(1.8,0);
    \draw[line width=1pt] (0,0)--(0,1.5);
    \draw[line width=1pt](1.8,0)--(1.8,1.5);
    \draw[line width=1pt](1.8,1.5)--(0,1.5);
    \draw[line width=1pt](0,0)--(-0.25,-0.75);
    \draw[line width=1pt](0,0)--(-0.75,-0.25);
    \draw[line width=1pt](1.8,0)--(2.8,-0.33);
    \draw[line width=1pt](1.8,0)--(2.13,-1);
    \draw[line width=1pt](0,1.5)--(-0.25,2.25) node [at end, above, blue] {\small $n$};
    \draw[line width=1pt](0,1.5)--(-0.75,1.75);
    \draw[line width=1pt](1.8,1.5)--(2.5,2) node [at end, above right, red] {\small 1};
    \draw[black,fill=gray] (0,0) circle (2.5ex);
    \draw[black,fill=lightgray] (1.8,0) circle (4.5ex);
    \draw[black,fill=white] (0,1.5) circle (2.5ex);
    \draw[black,fill=black] (1.8,1.5) circle (2.5ex);
    \node[black] at (1.8,0) {\scriptsize $k{-}2$};
    \draw[dotted] (-0.2,-0.6) to [out=-120,in=-160] (-0.6,-0.2);
    \node[below left] at (-0.25,-0.25) {\small $Q_2$};
    \draw[dotted] (2.05,-0.95) to [out=0,in=-60] (2.55,-0.28);
    \node[below right] at (2.3,-0.5) {\small $Q_1$};
    \draw[dotted] (-0.2,2) to [out=120,in=90] (-0.6,1.7);
    \node[above left] at (-0.25,1.75) {\small $Q_3$};
\end{tikzpicture}\hspace{-0.35cm}+...+\hspace{-0.35cm} \begin{tikzpicture}[scale=0.5,baseline={(0,0.375cm)}]
    \draw[line width=1pt] (0,0)--(1.8,0);
    \draw[line width=1pt] (0,0)--(0,1.5);
    \draw[line width=1pt](1.8,0)--(1.8,1.5);
    \draw[line width=1pt](1.8,1.5)--(0,1.5);
    \draw[line width=1pt](0,0)--(-1,-0.33);
    \draw[line width=1pt](0,0)--(-0.33,-1);
    \draw[line width=1pt](1.8,0)--(2.55,-0.25);
    \draw[line width=1pt](1.8,0)--(2.05,-0.75);
    \draw[line width=1pt](0,1.5)--(-0.25,2.25) node [at end, above, blue] {\small $n$};
    \draw[line width=1pt](0,1.5)--(-0.75,1.75);
    \draw[line width=1pt](1.8,1.5)--(2.5,2) node [at end, above right, red] {\small 1};
    \draw[black,fill=lightgray] (0,0) circle (4.5ex);
    \draw[black,fill=white] (1.8,0) circle (2.5ex);
    \draw[black,fill=white] (0,1.5) circle (2.5ex);
    \draw[black,fill=black] (1.8,1.5) circle (2.5ex);
    \node[black] at (0,0) {\scriptsize $k{-}1$};
    \draw[dotted] (-0.25,-0.95) to [out=0,in=-90] (-0.75,-0.28);
    \node[below left] at (-0.5,-0.5) {\small $Q_2$};
    \draw[dotted] (2,-0.6) to [out=0,in=-90] (2.4,-0.2);
    \node[below right] at (2.05,-0.25) {\small $Q_1$};
    \draw[dotted] (-0.2,2) to [out=120,in=90] (-0.6,1.7);
    \node[above left] at (-0.25,1.75) {\small $Q_3$};
\end{tikzpicture}\right)
\end{align}